\documentstyle[prl,aps,epsf,multicol,rotate]{revtex}
\title{Light Induced Melting of Colloidal Crystals in Two Dimensions}
\author{Erwin Frey$^1$, David R. Nelson$^1$, and Leo
  Radzihovsky$^2$} \address{$^1$Department of Physics, Harvard
  University, Cambridge, MA 02138} \address{$^2$Department of Physics,
  University of Colorado, Boulder, CO 80309}
\begin{document}
\bibliographystyle{prsty}
\date{\today}
\maketitle
\begin{abstract}
  We demonstrate that particles confined to two dimensions (2d) and
  subjected to a one-dimensional (1d) periodic potential exhibit a
  rich phase diagram, with both ``locked floating solids'' and smectic
  phases. The resulting phases and phase transitions are studied as a
  function of temperature and potential strength.  We find reentrant
  melting as a function of the potential strength. Our results lead to
  universal predictions consistent with recent experiments on 2d
  colloids in the presence of a laser-induced 1d periodic potential.
\end{abstract}
\vspace{-0.5cm}
\begin{multicols}{2}
\narrowtext 

Landau's order parameter expansion predicts that the direct transition
from a solid to a liquid should {\em always} be first
order~\cite{landau:37}.  In two dimensions (2d), however, fluctuations
can suppress the transition temperature so far below its mean-field
value that order parameter amplitude fluctuations (except in the form
of topological defects) play no role, and Landau's mean-field analysis
is qualitatively wrong. In this case a two-stage melting process
mediated by the unbinding of
dislocations~\cite{kosterlitz-thouless:73,nelson-halperin:79} and
disclinations~\cite{nelson-halperin:79} provides an alternative
scenario with two successive continuous phase transitions with an
intermediate hexatic phase instead of a single direct first order
transition.  A periodic embedding medium for the 2d solid (e.g. a
crystal substrate or a laser potential) leads to commensurability
effects and engenders an enormous variety of interesting
phenomena~\cite{pokrovsky-talapov-bak:86}.

Colloids confined between smooth glass plates provide an ideal model
system for experimental studies of 2d melting. In this system
individual particles can be imaged, allowing a direct observation of
topological defects and measurement of real-space correlation
functions.  Murray~{\em et\/al.\/}\cite{murray-sprenger-wenk:90} and
Zahn~{\em et\/al.\/}\cite{zahn-etal:99} have given strong experimental
evidence for a two-stage melting mechanism in such systems (because
dislocations relax slowly, long equilibration times are
required~\cite{murray-sprenger-wenk:90}).  Chowdhury~{\em
  et\/al.\/}\cite{chowdhury-ackerson-clark:85} studied the behavior of
a 2d colloidal suspension subjected to a 1d periodic potential
provided by the standing wave pattern of two interfering laser beams.
They observed a phenomenon called ``light induced freezing'', a
solidification driven by a commensurate laser potential.  Our research
is motivated by the recent work of Wei {\it
  et\/al.}~\cite{wei-bechinger-rudhardt-leiderer:98}, who
reinvestigated this phenomenon and discovered {\it reentrant melting}
upon gradually increasing the amplitude of the laser potential even
further.

Previous theoretical studies of this phenomenon have used density
functional theory~\cite{chakrabarti-etal:94} which gives essentially
the same results as Landau theory~\cite{chowdhury-ackerson-clark:85},
namely a change from first to second order critical behavior with
increasing potential strength.  Continuous melting is allowed because
the external potential singles out density modes in one direction and
the relevant Landau theory then contains only even powers in the local
Fourier components of the density $\rho_{\bf G} ({\bf r}) = \exp (i
{\bf G} \cdot {\bf r})$.  Unfortunately, the applicability of
mean-field-like theories to problems with continuous symmetry in 2d is
limited since they drastically underestimate the effect of
fluctuations.  Results from Monte-Carlo simulations are inconclusive:
Although earlier simulations~\cite{chakrabarti-etal:95} claimed to
have found a tricritical point at intermediate laser intensities and
reentrance, recent studies from the same
laboratory~\cite{das-sood-krishnamurthy:99} refute these results.  We
note that simulations with much larger numbers of particles have still
not completely resolved the nature of 2d melting {\em without} an
external potential~\cite{bagchi-andersen-swope:96}.

\begin{figure}[bht]
  \narrowtext \centerline{ \epsfxsize=0.521\columnwidth
  \epsfbox{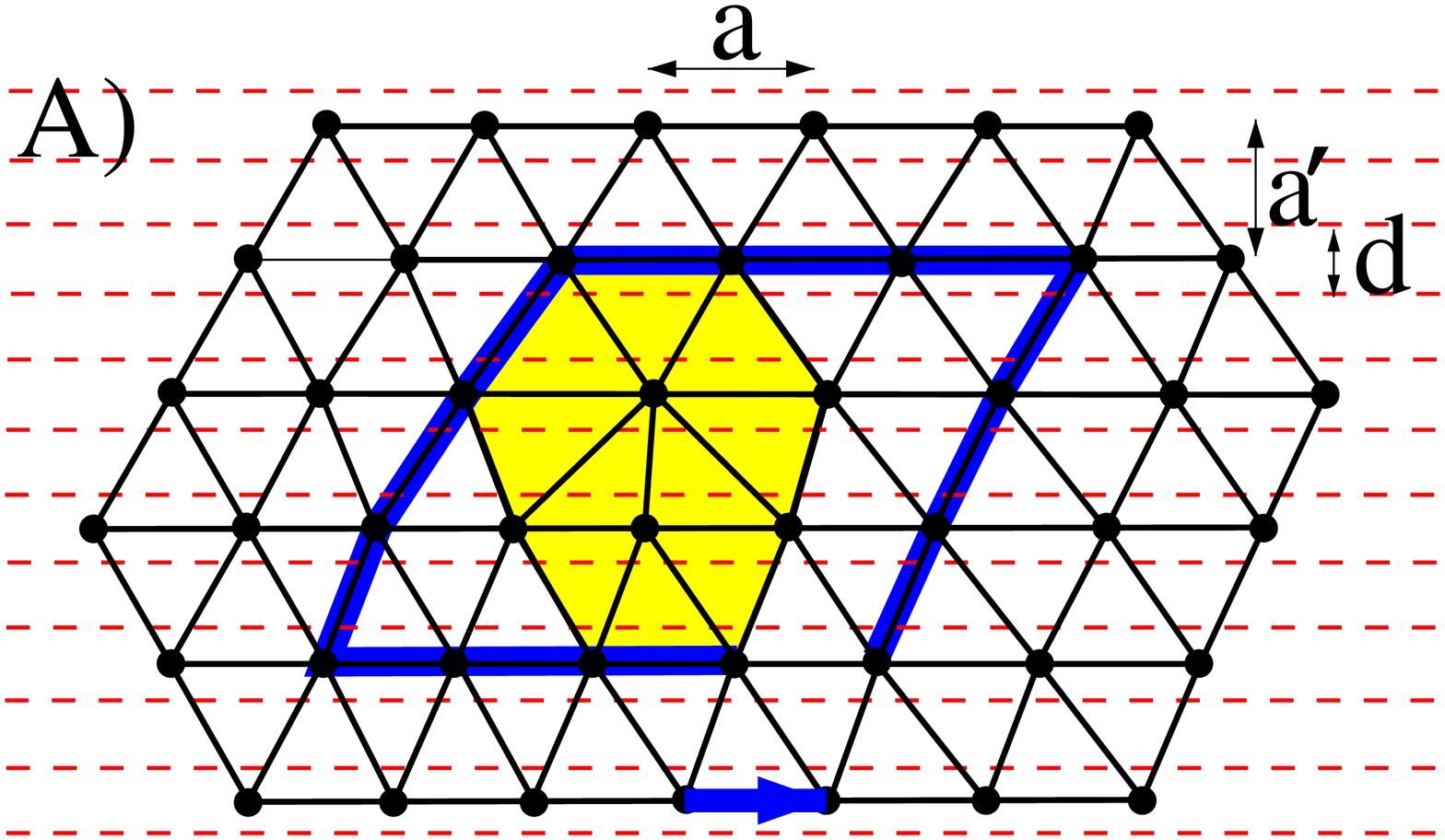} \hspace{0.02\columnwidth} \epsfxsize=0.45\columnwidth
  \epsfbox{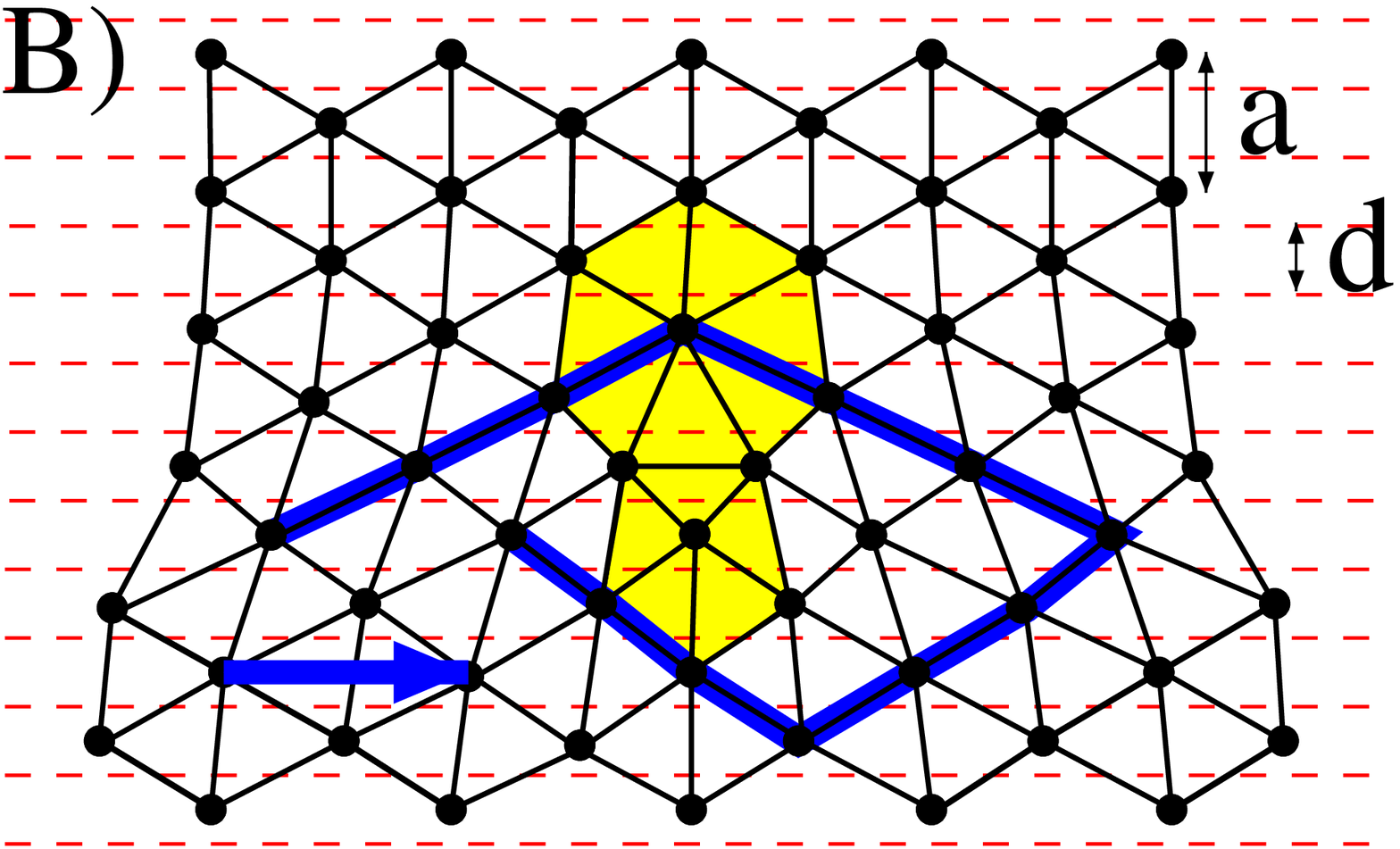}} \vspace{0.5truecm}
\caption{Triangular lattice with lattice constant $a$ subject to a
  periodic potential (maxima indicated by dashed lines) for two
  different relative orientations: A) $p_A d=a^\prime_A$ with
  $a^\prime_A = \sqrt{3}a/2$ and $p_A = 2$, and B) $p_B d=a^\prime_B$
  with $a^\prime_B = a/2$ and $p_B = 1$. Also shown are low energy
  dislocations with Burgers vector ${\bf b}$ parallel to the
  corrugation of the potential.}
\label{fig:dislocation}
\end{figure} 

In this Letter we take a different approach, building on concepts
developed in the context of dislocation mediated melting
theory~\cite{kosterlitz-thouless:73,nelson-halperin:79}. We model the
experimental system by a continuum elastic free energy $F = F_{\rm
  el}+F_{\rm p}$, where $F_{\rm el}= \frac12 \int d^2 r [2 \mu
u_{ij}^2 + \lambda u_{kk}^2]$ is the elastic energy associated with
the colloidal displacement field ${\bf u}(x,y)$ relative to the
equilibrium position in the unconstrained solid characterized by the
``bare'' elastic constants $\mu$ and $\lambda$. The quantity $u_{ij} =
\frac12 (\partial_i u_j + \partial_j u_i)$ is the 2d strain matrix.
The effect of the laser potential with troughs running along the
$x$-axis is described by $F_{\rm p} = \int d^2 r \, U (y,u_y)$, where
$U(y,u_y) = - U_0 a^{-2} \cos \left[ \frac{2\pi}{d} (u_y - \delta
  \cdot y ) \right]$. $U_0$ measures the strength of the laser
potential and the mean colloidal spacing $a$ is related to the
particle density $\rho$ by $\rho = 2/\sqrt{3} a^2$. A similar model
was used to discuss modulated superconducting films in
Ref.~\cite{pokrovsky-talapov-bak:86}.

Let $a'$ be the Bragg plane spacing for the orientation of the
colloidal crystal (relative to the troughs) which produces the lowest
free energy (see Fig.~\ref{fig:dislocation}).  Incommensurability
between $a$ and the distance between the troughs, $d$, is accounted
for by the mismatch parameter $\delta = a'/d - p$, where $p$ is the
integer closest to $a'/d$.

{F}or $\delta =0$, we can write $U(y,u_y) = - U_0 a^{-2} \cos [{\bf
  K}_p \cdot {\bf r}]$, where ${\bf K}_p = \frac{2 \pi}{a'} \, p \,
{\bf e}_y$ is a reciprocal lattice vector of the unperturbed system.
If $U_0=0$, there are algebraic Bragg peaks in the structure function
of the crystalline phase and $M_{{\bf K}} \equiv \langle \rho_{_{{\bf
      K}}} \rangle \sim 1/L^{{\overline \eta}_{{\bf K}}/2} \rightarrow
0$ as the system size $L\rightarrow \infty$, where ${\overline
  \eta}_{{\bf K}}= \frac{k_B T}{4 \pi} \frac{3 \mu + \lambda}{\mu (2
  \mu + \lambda)} {\bf K}^2$~\cite{nelson-halperin:79}. It is
straightforward to show that $M_{{\bf K}}$ then vanishes for small
$U_0$ as $M_{{\bf K}} \sim \mid U_0 \mid^{{\overline \eta}_{{\bf
      K}}/(4-{\overline \eta}_{{\bf K}})}$\cite{note1}. In contrast,
$M_{{\bf K}}$ should always vanish {\em linearly} with $U_0$ in the liquid
and hexatic phases \cite{murray-sprenger-wenk:90,zahn-etal:99} of the
unperturbed colloid. The laser potential will also induce long range
bond orientational order in $\psi_6 = \langle e^{6i\theta({\bf r})}
\rangle$~\cite{balents-nelson:94}. The bond order parameter $\psi_6$
vanishes linearly with $U_0$ in the liquid, vanishes like a power of
$U_0$ in the hexatic phase ($\psi_6 \sim \mid U_0 \mid^{6{\overline
    \eta}_{6}/(4-{\overline \eta}_{6})}$, where ${\overline \eta}_6$
is the exponent describing the algebraic decay of bond order), and
approaches a nonzero constant as $U_0 \rightarrow 0$ in the solid
phase.

\begin{figure}[thb]
  \narrowtext \centerline{\epsfxsize=0.47\columnwidth
    \epsfbox{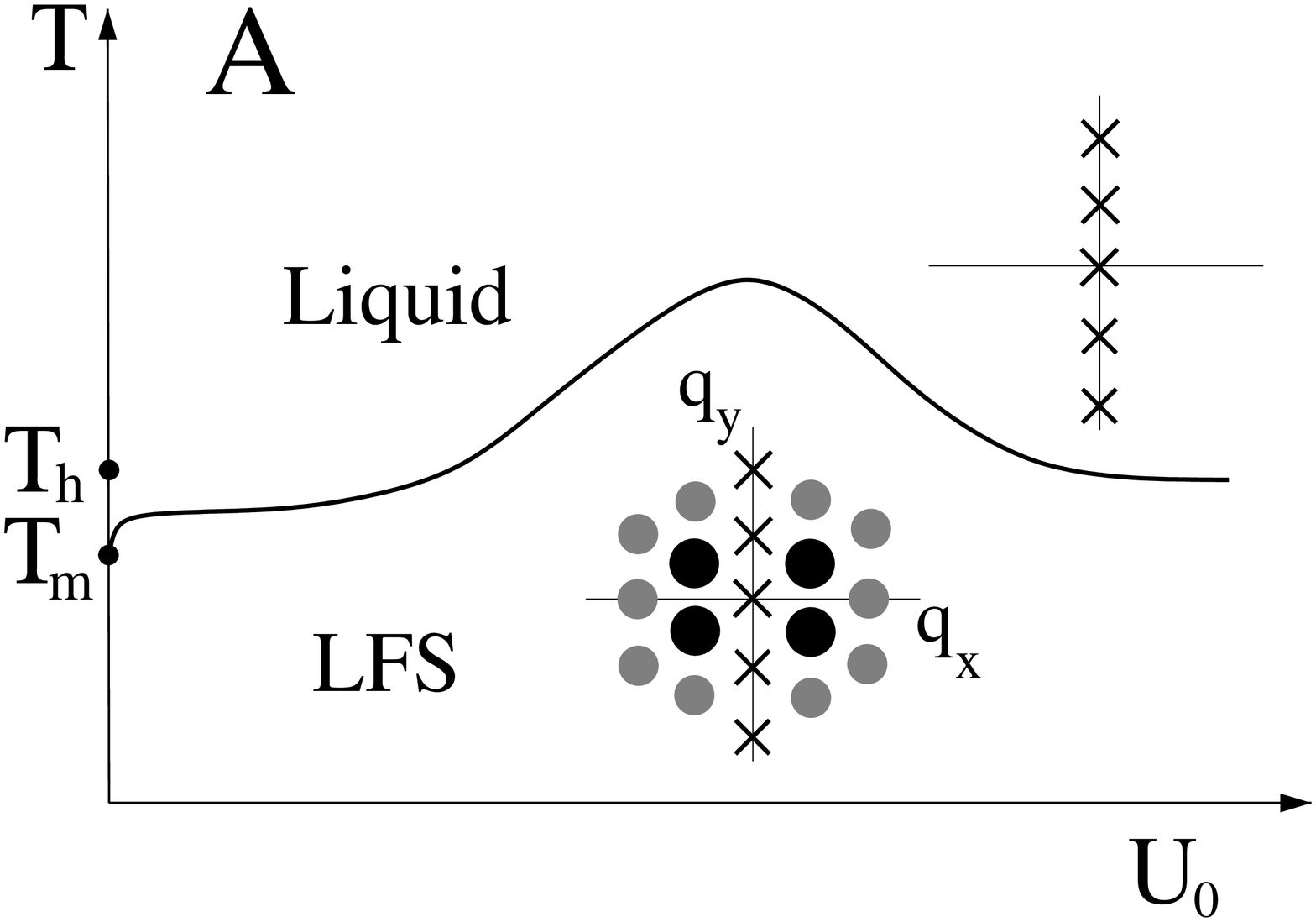} \hspace{0.04\columnwidth}
    \epsfxsize=0.47\columnwidth \epsfbox{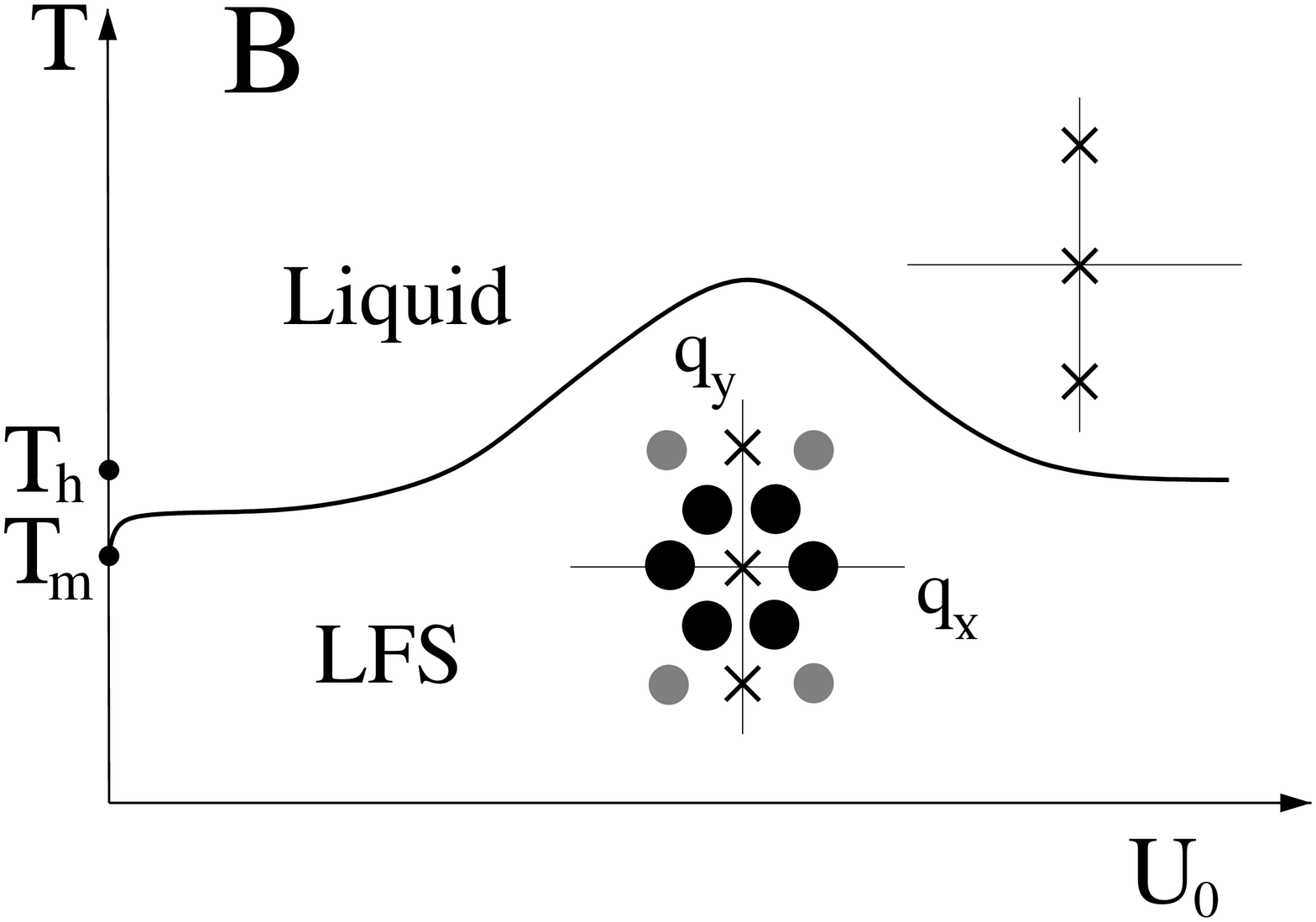}} \vspace{0.5cm}
\caption{Schematic $p=1$ phase diagram for orientations A and B. 
  {\em Insets:} Schematic structure functions in the various phases.
  The $\times$'s indicates delta-function Bragg peaks and circles
  algebraic peaks.}
\label{fig:phase_diagram_small}
\end{figure}

In our analysis of larger values of $U_0$, we shall also focus on the
commensurate case ($\delta = 0$), where the spacing $a^\prime$ of the
lattice planes parallel to the troughs equals an integer $p$ times
$d$. In Fig.~\ref{fig:dislocation}, two particularly interesting
orientations, denoted A and B, of the colloidal crystal and
laser potential are shown.  Only dislocations with Burgers parallel to
the troughs have the usual logarithmically divergent energy.  In
orientation A, four of the six fundamental Burgers vectors are
disfavored by the potential, which requires that they be attached to a
semi-infinite discommensuration
string~\cite{pokrovsky-talapov-bak:86}. In orientation B, all {\em
  six} of the fundamental Burgers vectors are disfavored. The lowest
energy Burgers vector parallel to the troughs has length $\sqrt{3} a$.
At sufficiently low temperatures, the laser potential is always
relevant~\cite{nelson-halperin:79}, leaving massless phonon
displacements $u_x$ along the troughs with massive out-of-valley
displacements $u_y$.  We call this ordered phase a {\em ``locked
  floating solid''} (LFS), reflecting its resistance to strains
associated with $u_y$, and its ability to accomodate strains in $u_x$.
In reciprocal space, the LFS is characterized by a structure function
$S({\bf q})$ with a row of delta-function Bragg peaks at $G_y^{(n)} =
2 \pi n / d$ ($n \in Z$) along the $q_y$-axis and power law Bragg
peaks off this axis (see Fig.~\ref{fig:phase_diagram_small}).

Upon integrating out the massive $u_y$-modes and using standard
renormalization group methods
\cite{nelson-halperin:79,radzihovsky-frey-nelson:99} to eliminate
bound dislocation pairs in the LFS phase, we are left with a free
energy with temperature and potential strength dependent {\em
  effective} elastic constants,
\begin{eqnarray}
  F_{\text{LFS}} = \frac{1}{2} \int d^2 r \left\{
  K_{\text{eff}} \, (\partial_x u_x)^2 +
  \mu_{\text{eff}} \, (\partial_y u_x)^2 \right\} \, .
\label{eq:KT_free_energy}
\end{eqnarray}
Because only dislocations with Burgers vectors parallel to the troughs
are important, the physics is described by an anisotropic scalar
Coulomb gas or, equivalently, an anisotropic 2d XY model. For $p$
smaller than a critical value $p_c$ we find that the LFS melts with
increasing temperature at
\begin{eqnarray}
  k_B T_m =   \frac{\sqrt{K_{\text{eff}} \, 
          \mu_{\text{eff}}}}{8 \pi} \, b^2 \, , 
\label{eq:KT_melting_temperature}
\end{eqnarray}
via an unbinding of dislocations with Burgers vectors ${\bf b}
\parallel {\bf e}_x$ {\em before} the laser potential which gives the
mass to the $u_y$-modes become irrelevant (see below).  For $T<T_m$
the correlation function $C_{\bf G} ({\bf r})= \langle
\rho^{\phantom{*}}_{\bf G} ({\bf r}) \rho^*_{\bf G} (0) \rangle$ for
reciprocal lattice vectors with $G_x \neq 0$ shows power law decay
$C_{\bf G} ({\bf r}) \sim \mid (\mu_{\rm eff}/K_{\rm eff})^{1/2} x^2 +
(K_{\rm eff}/\mu_{\rm eff})^{1/2} y^2\mid^{-\eta_{\bf G}/2}$. The
structure function is singular, $S({\bf q}) \sim 1/ \mid {\bf q} -
{\bf G} \mid^{2-\eta_{\bf G}}$, near the reciprocal lattice vector
${\bf G}$~\cite{nelson-halperin:79}.  The exponents $\{ \eta_{\bf G}
\}$ are given in terms of the elastic constants by $\eta_{\bf G} = k_B
T G_x^2 / 2 \pi \sqrt{K_{\rm eff} \mu_{\rm eff}}$.

Unlike conventional 2d melting~\cite{nelson-halperin:79}, $\eta_{\bf
  G}$ is {\em universal} at the melting transition, and given by
$\eta_{\bf G}^* = ({\bf G}\cdot{\bf b}/4 \pi)^2$, where ${\bf b}$ is
the smallest allowed Burgers vector in the trough
direction~\cite{radzihovsky-frey-nelson:99}.  For orientation A, $b=a$
and the exponent characterizing the algebraic order in the off-axis
peaks (see insets of Fig.~\ref{fig:phase_diagram_small}) closest to
the $q_y$-axis is $\eta_{\bf G}^* = 1/4$; for the next row of peaks
with $G_x = 4 \pi / a$ one gets $\eta_{\bf G}^* = 1$, consistent with
the algebraic decay observed in
Ref.~\cite{wei-bechinger-rudhardt-leiderer:98}. For orientation B,
$b=\sqrt{3} a$ and the six quasi Bragg peaks closest to the origin
have {\em different} power laws; peaks on the x-axis have $\eta_{\bf
  G}^* =1$, whereas the four off-axis peaks have $\eta_{\bf G}^*=1/4$.
Just above the melting temperature, we expect diverging translational
correlation lengths (defined by the widths of Lorentzian peaks in the
structure function) parallel and perpendicular to the troughs $\xi_x
\sim \xi_y \sim \exp [\text{const.}/\mid T-T_m\mid^{1/2}]$.

To obtain the melting curve as a function of $U_0$ we calculate
$K_{\rm eff}$ and $\mu_{\rm eff}$ and use
Eq.~(\ref{eq:KT_melting_temperature}).  We start from a microscopic
model with a screened repulsive Coulomb interaction $V(r) = V_0 a \,
\exp (- \kappa r) / r$, where the screening length $\kappa^{-1}$ is
typically much smaller than $a$ and $V_0$ depends on the dielectric
constant, $\kappa$ and the sphere
radius~\cite{wei-bechinger-rudhardt-leiderer:98}. In such a dilute
limit the two Lam{\'e} coefficients are equal, $\lambda = \mu$.  Hence
the Kosterlitz-Thouless melting temperature in the absence of a laser
potential ($U_0=0$) is simply given by~\cite{nelson-halperin:79} $k_B
T_m^0 = \mu a^2 / 6 \pi$.  In the opposite limit of {\em infinite}
potential strength the effective free energy simplifies to
Eq.~(\ref{eq:KT_free_energy}) with $K_{\rm eff} = 3 \mu$ and $\mu_{\rm
  eff} = \mu$.  Using Eq.~(\ref{eq:KT_melting_temperature}) this gives
$k_B T_m^\infty = \sqrt{3} \mu a^2 / 8 \pi \approx 1.3 \, k_B T_m^0$.
One might have thought that the melting temperature simply increases
monotonically with $U_0$ from $T_m^0$ to $T_m^\infty$.  It turns out,
however, that fluctuations in the $u_y$-modes caused by lowering the
potential strength from infinity typically {\em increase} the melting
temperature.  To see this, we integrate out the $u_y$-modes using the
screened Coulomb potential; to leading order in $k_B T / U_0$ and $V_0
e^{- \kappa a} / U_0$ we find (for orientation
A)~\cite{radzihovsky-frey-nelson:99}
\begin{eqnarray}
  \mu_{\rm eff} &\approx& \mu 
  \left\{ 1 + \frac{9 (\kappa a)^2}{64 \pi^2} 
              \, \left( 1+ \frac{17}{3 \kappa a} \right) 
              \, \frac{k_B T}{p^2 U_0}
  \right\} \, ,\\
  K_{\rm eff} &\approx& K 
  \left\{ 1 + \frac{(\kappa a)^2}{64 \pi^2} 
              \left( 1-8v-\frac{23+104v}{3\kappa a}  
              \right) \frac{k_B T}{p^2 U_0}
  \right\} \, ,
\end{eqnarray}
where $v=V_0 e^{-\kappa a} / k_B T$, $\mu = \frac38 v k_BT \kappa^2$
and $K=3 \mu$. Lowering the potential strength $U_0$ always increases
the shear modulus, whereas the behavior of the compressional modulus
depends on the magnitude of $v$ and $\kappa a$. When combined with
Eq.~(\ref{eq:KT_melting_temperature}), these expressions imply that
the melting temperature $T_m$ increases with decreasing $U_0$ for
$\kappa a \gtrsim 5.6$ (in
Ref.~\cite{wei-bechinger-rudhardt-leiderer:98} $\kappa a \approx 10$),
\begin{eqnarray}
  T_m = T_m^\infty 
                 \left\{ 
                       1 + \frac{5[(\kappa a)^2\!-\!31]}{64 \pi^2}  
                           \left(1\!+\! \frac{13}{3 \kappa a} \right) 
                           \frac{k_B T_m^\infty}{p^2 U_0} 
                 \right\} \, ,
\end{eqnarray}
thus implying reentrant melting for a band of temperatures as a
function of potential strength (see
Fig.~\ref{fig:phase_diagram_small}).  For a more precise estimate of
the phase boundary one would also need to determine the
renormalization of the effective elastic constants by phonon
nonlinearities and by bound dislocation pairs.  In general, however,
one expects only small downward renormalizations with increasing
temperature~\cite{fisher:82} which will not affect the existence of
reentrant melting.For small $U_0$, we find that the melting curve has
a universal shape $T_m (U_0) -T_m(0) \sim [\ln (k_B T_m /
U_0)]^{-1/{\overline \nu}}$ with ${\overline \nu} \approx 0.36963
\cdots$.

\begin{figure}
  \narrowtext \centerline{ \epsfxsize=0.47\columnwidth
    \epsfbox{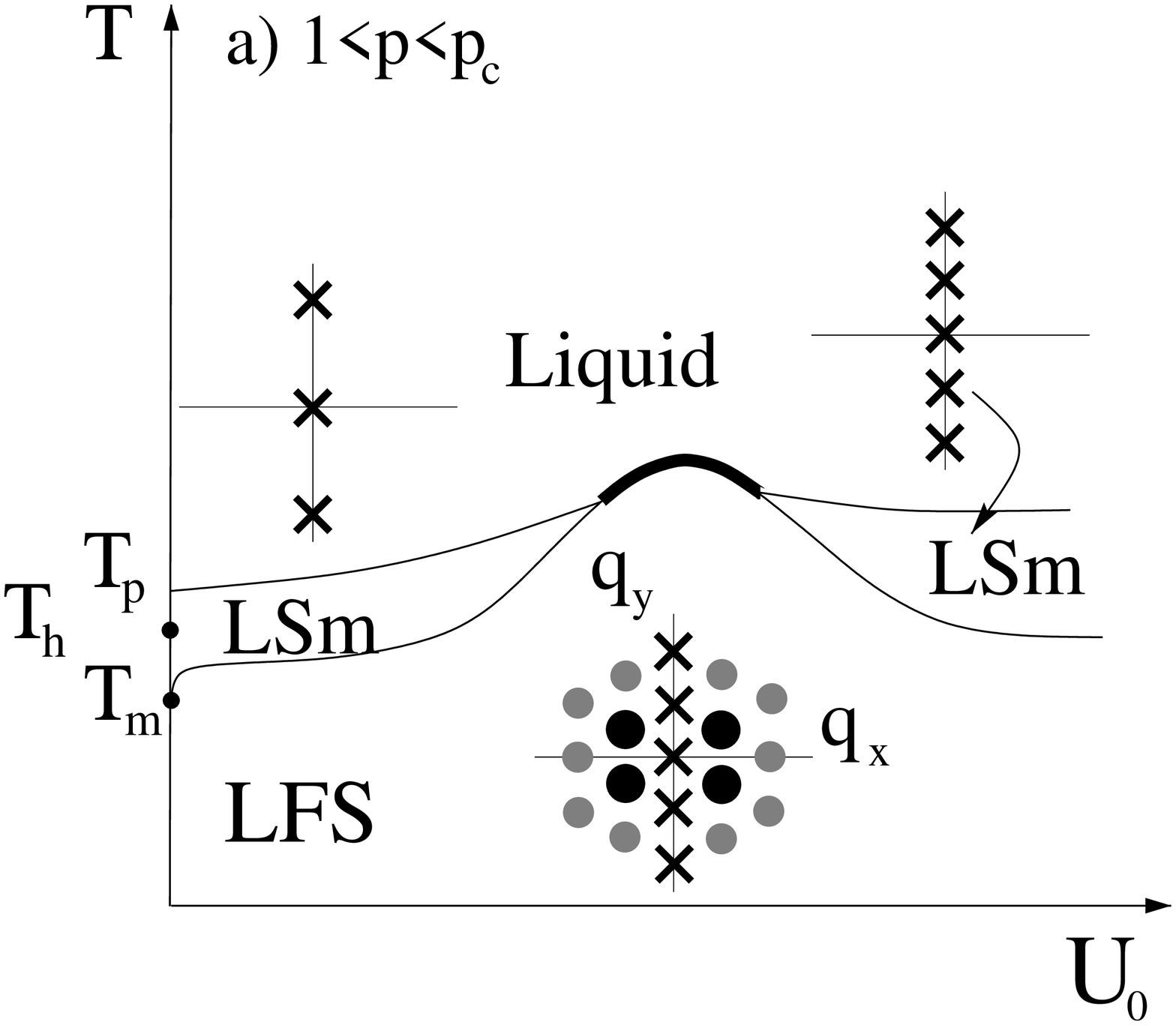} \hspace{0.02\columnwidth}
    \epsfxsize=0.47\columnwidth \epsfbox{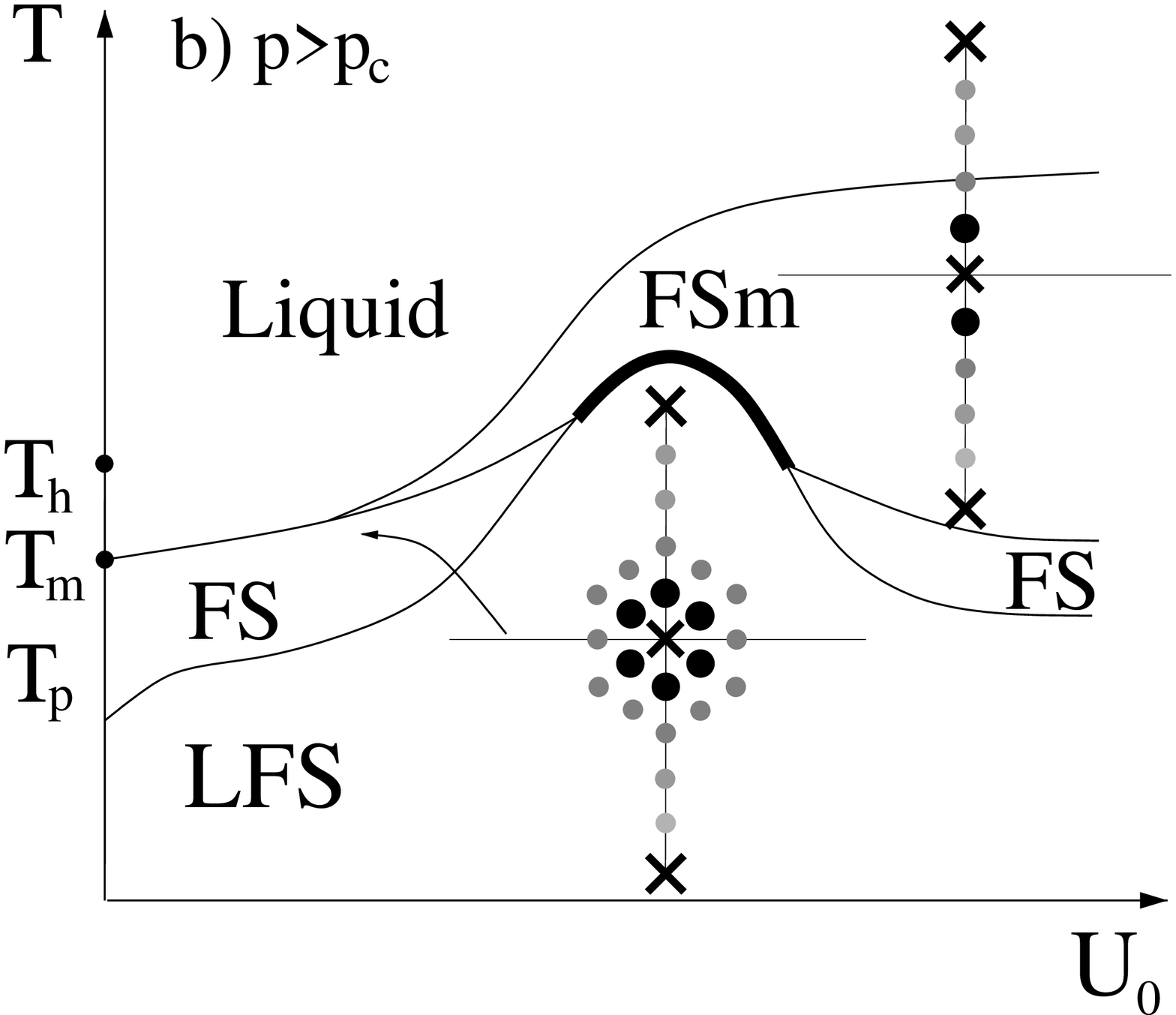}}
\vspace{0.5cm}
\caption{Schematic phase diagrams for a) $1<p<p_c$ and 
  b) $p > p_c' > p_c$ (for orientation A).  Heavy lines indicate phase
  transitions which are most likely of first order. {\em Insets:}
  Schematic structure functions.}
\label{fig:phase_diagram_large}
\end{figure}

We now discuss the topology of the phase diagram for larger values of
the integer commensurability parameter $p$.  As illustrated in
Fig.~\ref{fig:phase_diagram_large}, for $p>p_c$ a floating solid (FS)
with {\em two} soft phonon modes \cite{nelson-halperin:79} can
intervene (barring a direct first order transition) between the LFS
and the liquid phase. To determine $p_c$ one must calculate the
thermal renormalization of the laser potential, defined by $ U_0 (s) =
U_0 s^2 \exp\left[ - \frac{1}{2} K_p^2 \langle u_y^2 \rangle_> \right]
\equiv U_0 s^{\lambda_p}$, where $\lambda_p$ is a renormalization
group eigenvalue and the subscript ``$>$'' indicates that the
$u_y$-modes are integrated over a momentum shell
$[\Lambda/s,\Lambda]$. The laser potential becomes irrelevant at long
wavelengths whenever $\lambda_p<0$. The thermal average over the
$u_y$-modes requires an effective elastic long wavelength free energy
appropriate to the FS phase.  Because the laser potential breaks
rotational symmetry, this coarse-grained free energy contains {\em
  six} independent elastic constants,
\begin{eqnarray}
F_{\rm FS} &=& \int d^2 r \,
  \bigl\{ 2 \mu u_{xy}^2 + \frac12 \lambda_{xx} u_{xx}^2 
      + \frac12 \lambda_{yy} u_{yy}^2 \nonumber \\
      && \qquad + \lambda_{xy} u_{xx} u_{yy} 
      + 2 \gamma \theta^2 + 2 \alpha \theta u_{xy}
 \bigr\} \, ,
\label{eq:free_energy_uniaxial}
\end{eqnarray}
where $\theta = \frac{1}{2} (\partial_x u_y - \partial_y u_x)$ is the
local rotation angle induced by the phonon displacements.  The
resulting expression for $\lambda_p$ is rather complicated, with an
implicit dependence on the strength of the laser potential and the
temperature~\cite{radzihovsky-frey-nelson:99}. However, a useful
estimate results from neglecting the effect of the substrate on the
elastic coefficients and considering an isotropic elastic free energy.
For orientation A, one finds~\cite{nelson-halperin:79} $\lambda_p = 2
- \frac{4 \pi}{3} \, \frac{k_B T}{{\bar \mu} a^2} \, p^2$ with the
effective elastic constant ${\bar \mu} = 2 \mu (2 \mu + \lambda) / (3
\mu + \lambda)$.  In the dilute limit $\kappa a \gg 1$, relevant for
many colloidal systems, we have $\mu = \lambda$ and the eigenvalue is
negative for $T>T_p$, with ${\bar \mu} = 3 \mu / 2$ and $ k_B
T_p^{\rm dil} \approx 9 \mu a^2 / 4 \pi p^2$.  When compared with the
melting temperature of the LFS phase in the same limit of $\kappa a
\gg 1$ for weak potentials, $k_B T_m^0 \approx \mu a^2 / 6 \pi$, we
find that a window of the FS phase exists for $p>p_c \approx 3
\sqrt{3/2} \approx 3.7$ for orientation A~\cite{note2}.  Note,
however, that $p_c$ is not a universal constant but implicitly depends
on the strength of the laser potential, $p_c (U_0)$.

{F}or intermediate commensurate densities $1<p<p_c$, no FS phase
exists, and the dislocation unbinding transition discussed above melts
the LFS to a locked smectic (LSm) phase.  In the LSm phase the laser
potential is relevant and only one out of $p$ possible troughs is
preferentially occupied by the colloidal particles. At higher
temperatures an Ising transition ($p=2$) (more generally, $p$-state
clock model transition~\cite{jose-etal:77}) takes the system to a
modulated liquid phase.  We expect the same kind of reentrant behavior
for the LFS-LSm phase boundary as discussed above for $p=1$. However,
the LSm-Liquid phase boundary should depend more weakly on the
potential strength: since the shear modulus is zero in the LSm phase
the reentrance mechanism discussed above does not apply here.

For $p>p_c$ the laser potential becomes irrelevant {\em before} the
LFS melts (via a roughening-like transition), leading to the
uniaxially anisotropic FS described by the elastic free energy $F_{\rm
  FS}$, Eq.~(\ref{eq:free_energy_uniaxial}).  Melting of this uniaxial
solid (with $\gamma=\alpha=0$) has been studied in detail by Ostlund
and Halperin~\cite{ostlund-halperin:81}. Due to the uniaxial
anisotropy there are two different types of dislocations: type-I
dislocations with Burgers vector along a reflection symmetry axis of
the solid, and type-II dislocations at angles $\pm \phi_0$ with
respect to this axis.  The laser potential in
Fig.~\ref{fig:dislocation}A is oriented in such a way that type-I
dislocations are parallel to the troughs of the laser potential,
interact more weakly, and will therefore unbind first. This ``type-I''
melting transition leads to a ``floating smectic'' phase (FSm)
stabilized by the laser potential such that it retains quasi long
range order in $\rho_{\bf G} = \exp[ i {\bf G} \cdot {\bf u}]$ for
${\bf G} = 2 \pi {\bf e}_y / pd$. We expect that a FSm phase
intervenes between a FS and the liquid if $p$ is sufficiently large
($p \geq p_c^\prime = 4$~\cite{jose-etal:77}).
The novelty of this FS-FSm type-I melting is that, in contrast to the
similar melting of the LFS, the destruction of translational order in
$u_x$ by dislocations takes place in the presence of a coupled
``spectator'' massless phonon mode $u_y$.  To analyze this transition
we proceed in the standard way~\cite{nelson-halperin:79,jose-etal:77}
by introducing type-I dislocations into the elastic free energy and
performing a duality transformation on the resulting Coulomb gas
Hamiltonian to convert it into a modified Sine-Gordon model
\begin{eqnarray}
  H = \int d^2 r \left\{ 
      \frac{|{\bbox\nabla} h|^2}{2 K_I} 
    - g \cos{[a(h+i\psi)]} 
                 \right\} 
    + F_{FS}[{\bbox \phi}] \, .
\end{eqnarray}
Here $h (x,y)$ is a dummy field which when integrated out in $\exp
(-H/k_B T)$ gives rise to a scalar Coulomb gas with interaction
strength proportional to $K_I$. The quantity ${\bbox \phi}$ is the
single valued part of the displacement fields, $\psi({\bf
  q})=(\lambda_{xy} q_y^2 - (\mu - \gamma) q_x^2) u_y({\bf q}) / q^2$
and $g\equiv 2e^{-E_c/k_B T}$ with $E_c$ the core energy of type-I
dislocations. The melting temperature can be determined from the
condition that the renormalization group eigenvalue of $g$ vanishes,
\begin{eqnarray}
  k_B T_{\rm FS-FSm} = 
 \frac{a^2}{8 \pi} \left( \sqrt{\lambda_{xx} \mu_{-}}
  - c \frac{\mu^2}{\sqrt{\mu_{+} \lambda_{yy}}}  \right) \, ,
\end{eqnarray}
with suitable renormalized elastic constants. Here $\mu_\pm =
\mu+\gamma\pm \alpha$ and $c$ is a dimensionless function of ratios of
the FS elastic constants~\cite{radzihovsky-frey-nelson:99}.  The
effect of the ``spectator'' phonon modes is to reduce the melting
temperature.  Just below $T_{\rm FS-FSm}$ we find for the renormalized
stiffness $K_I^R(T) \approx K_I^R(T_{\rm FS-FSm}) [1 + {\rm const.}
(T_{\rm FS-FSm} - T)^{1/2}]$. The coupling between $u_x$ also induces
cusps in the elastic constants of the spectator field $u_y$. Above the
melting temperature $K_I$ is zero, type-I dislocations are unbound,
and at long length scales the elastic free energy describing the
floating smectic phase is of the form $ F_{\rm FSm} = \frac12 \int d^2
r \left\{ B_{x} (\partial_x u_y)^2 + B_{y} (\partial_y u_y)^2
\right\}$ with $B_x \approx [4\gamma \mu -\alpha^2] / (\mu + \gamma -
\alpha)$ and $B_y \approx \lambda_{yy} - \lambda_{xy}^2/\lambda_{xx}$.
Note that $B_x$ vanishes in the limit $\gamma = \alpha = 0$ as one
would expect for a rotationally invariant 2d smectic.  The structure
factor in the FSm phase exhibits power-law singularities at reciprocal
lattice vectors ${\bf G}_n = 2 \pi n {\bf e}_y / pd$ (for $n \neq p$),
with an exponent $\eta_{\rm FSm} = {k_B T \, G_n^2}/2 \pi \sqrt{B_x
  B_y}$. The FSm phase melts via a second dislocation unbinding
transition when $\eta_{\rm FSm} \geq 1/4$.

We have only discussed laser potentials which are commensurate with
the colloidal crystal. Sufficiently large incommensurability favors
the discommensurations attached to Burgers vectors with an unfavorable
orientation relative to the troughs \cite{pokrovsky-talapov-bak:86},
and leads to an enormous variety of phase transitions as a function of
density at fixed trough spacing which we hope to discuss in a future
publication.

It is a pleasure to thank B.I. Halperin for helpful discussions, and
C. Bechinger and P.  Leiderer for communicating unpublished results.
This work has been supported by the DFG through a Heisenberg
fellowship (Fr~850/3-1) (ef), by the NSF through Grant No.
DMR-9714725 (drn), Harvard's MRSEC Grant No.~DMR98-09363 (drn), Grant
No.~DMR-9625111 (lr), University of Colorado's MRSEC Grant
No.~DMR-9809555 (lr), and by the A.P. Sloan and the Packard
Foundations (lr).

\end{multicols}
\end{document}